\documentclass[aip,jcp,reprint,twocolumn,superscriptaddress,showpacs,floatfix]{revtex4-2}
\usepackage{amsmath}
\usepackage{microtype}
\usepackage{graphicx}% Include figure files
\usepackage{dcolumn}% Align table columns on decimal point
\usepackage{bm}% bold math
\usepackage{verbatim}
\usepackage[utf8]{inputenc}
\usepackage[T1]{fontenc}
\usepackage{mathptmx}
\usepackage{etoolbox}
\usepackage{subcaption} % For subfigure support
\usepackage{float}
\usepackage{mathrsfs}
\usepackage{dcolumn}
\usepackage{multirow}
\usepackage{ragged2e}
\DeclareUnicodeCharacter{2212}{-}

\usepackage{balance}
\usepackage{caption}
\usepackage{hyperref}
\hypersetup{
    colorlinks=true,    % Enables colored links
    linkcolor=blue,     % Color of internal links
    citecolor=magenta,      % Color of citations
    pdfborder={0 0 0}   % Removes the red boxes around links
}

\usepackage{booktabs}

\makeatletter
\def\@email#1#2{%
 \endgroup
 \patchcmd{\titleblock@produce}
  {\frontmatter@RRAPformat}
  {\frontmatter@RRAPformat{\produce@RRAP{*#1\href{mailto:#2}{#2}}}\frontmatter@RRAPformat}
  {}{}
}%
\makeatother

\begin{document}
\preprint{AIP/123-QED}

\title{Inclusion of Three-body Corrections in the Relativistic Equation-of-Motion Coupled Cluster Method: Application to Electron Detachment}
% Force line breaks with \\
\author{Mrinal Thapa}
\affiliation{\small Department of Chemistry, Indian Institute of Technology Bombay, Powai, Mumbai 400076, India}

\author{Achintya Kumar Dutta}
%%\thanks{Corresponding author}
\email[e-mail: ]{achintya@chem.iitb.ac.in}
\affiliation{\small Department of Chemistry, 
Indian Institute of Technology Bombay, Powai, Mumbai 400076, India}
\affiliation{ \small Department of Inorganic Chemistry, Faculty of Natural Sciences, Comenius University, Ilkovičova 6, Mlynská dolina 84215 Bratislava, Slovakia \\
}%
\email[e-mail: ]{achintya.kumar.dutta@uniba.sk}

\date{\today}% It is always \today, today,
             %  but any date may be explicitly specified

\begin{abstract}
We present the formulation and implementation of a triples-correction scheme for relativistic equation-of-motion coupled-cluster calculations of ionization potentials. Both full and partial triples-correction schemes are implemented using the exact two-component atomic mean-field (X2CAMF) Hamiltonian in combination with Cholesky decomposition (CD) of two-electron integrals and a frozen natural spinor (FNS) truncation scheme to reduce computational cost. Benchmark calculations on halide anions, noble gas atoms, hydrogen halides, and dihalogen molecules demonstrate that triple excitations are essential for quantitative ionization potentials, reducing mean absolute errors to approximately 0.01–0.08 eV relative to reference and experimental values. The X2CAMF approximation reproduces four-component Dirac–Coulomb results with negligible deviations, while the CD-FNS strategy substantially reduces computational costs. The resulting partial triples-correction scheme scales as $\mathcal{O}(n^7)$ with storage comparable to CCSD, providing an accurate and practical approach for relativistic ionization-energy calculations in heavy-element systems. 
\end{abstract}

\maketitle
 \begin{comment}
\begin{quotation}
The ``lead paragraph'' is encapsulated with the \LaTeX\ 
\verb+quotation+ environment and is formatted as a single paragraph before the first section heading. 
(The \verb+quotation+ environment reverts to its usual meaning after the first sectioning command.) 
Note that numbered references are allowed in the lead paragraph.
%
The lead paragraph will only be found in an article being prepared for the journal \textit{Chaos}.
\end{quotation}
\end{comment}

\section{\label{sec:level1}Introduction}
Ionization is a fundamental process with far-reaching applications in areas such as spectroscopy, chemical reactivity, and plasma physics. Reliable theoretical models of ionized states are essential for interpreting experimental data. A wide variety of \textit{ab initio} quantum chemical methods are available for simulating the ionization of atoms and molecules. Among these methods, equation-of-motion coupled-cluster (EOM-CC) theory stands out because of its systematically improvable nature and the capacity to simulate multiple ionized states in a single calculation.\cite{cizekCorrelationProblemAtomic1966, CCSD1972,  stantonEquationMotionCoupledcluster1993} The EOM-CC method yields ionization potentials (IP) identical to those obtained from the linear response coupled-cluster (CC) approach, although the two approaches are derived from very different theoretical perspectives.\cite{mukherjeeResponsefunctionApproachDirect1979, monkhorstCalculationPropertiesCoupledcluster1977, kochCoupledClusterResponse1990} An alternative approach for calculating ionization energies in the CC framework is the $\Delta$CC method, in which the IP is obtained as the energy difference between two separate ground-state calculations.\cite{bagusSelfConsistentFieldWaveFunctions1965,bagusDirectNearHartreeFockCalculations1971} The EOM-CC approach offers advantages over the $\Delta$CC approach because it can compute multiple ionized states within a single calculation and also provides access to transition properties. The equation-of-motion CC formalism for IP (IP-EOM-CC) calculations has also been extended to the relativistic regime\cite{pathakip, fns_ip_2022,CD_IP_2025,Chengcore,SheeIP} using both two-component (2c) and four-component (4c) relativistic Hamiltonians. In practice, relativistic IP-EOM-CC calculations are usually performed in the singles and doubles approximation (IP-EOM-CCSD), which scales as $\mathcal{O}(n^6)$, while the EOM part scales as $\mathcal{O}(n^5)$. To achieve quantitative accuracy in IP, three-body operators must be included. Musia{\l} and co-workers have described the full inclusion of three-body operators in the IP-EOM-CC method.\cite{10.1063/1.1527013} To the best of our knowledge, the only study reporting the inclusion of full triple excitations in relativistic EOM-CC methods for IP is that of DePrince and co-workers.\cite{deprinceip} Inclusion of partial and full triples corrections has been reported for the related (0,1) sector of the relativistic Fock-space multi-reference CC method.\cite{sym12071101,fsccsdt}

The full inclusion of triples in the relativistic IP-EOM-CC method (IP-EOM-CCSDT) scales as $\mathcal{O}(n^8)$, with an $\mathcal{O}(n^6)$ storage requirement, due to the need for storing ground-state triples amplitudes. The computational cost of relativistic IP-EOM-CCSDT is much higher than the corresponding non-relativistic version, owing to the lack of spin symmetry, the use of uncontracted basis sets, and the need to store complex-valued matrix  elements.\cite{reiherRelativisticQuantumChemistry2015, dyallIntroductionRelativisticQuantum2007} Consequently, its application is restricted to small basis sets. Various strategies can be used to reduce the computational cost of relativistic IP-EOM-CCSDT calculations. The formal scaling can be reduced by perturbative truncation of the three-body operator. The number of floating-point operations can be reduced by using truncated natural spinors.\cite{fns_ccsd_2022, Visccher_2022} The efficiency of frozen natural spinors can be further improved by combining them with the Cholesky decomposition (CD) based implementation of two-electron integrals.\cite{CD_CCSD_2025, CD_IP_2025} The aim of this manuscript is to develop an efficient triples correction scheme for the relativistic EOM-CC method for IP.

\section{\label{sec:level2}Theory and Computational Details}
\subsection{Relativistic EOM-CC for Ionized States}
One of the most straightforward ways to include relativistic effects in quantum chemical calculations is to use the 4c Dirac--Coulomb (DC) Hamiltonian,
\begin{equation}
\label{eqn1}
    {{\hat{H}}^{\text{4c}}}=\sum\limits_{i}^{N}{\left[ c{{{\vec{\alpha }}}_{i}}.{{{\vec{p}}}_{i}}+{{\beta }_{i}}{{m}_{0}}{{c}^{2}}+\sum\limits_{A}^{{{N}_{\text{nuc}}}}{{{{\hat{V}}}_{iA}}} \right]}+\sum\limits_{i<j}^{N}{\frac{1}{{{r}_{ij}}}}
\end{equation}
Here, $\alpha$ and $\beta$ denote the Dirac matrices, and $\hat{p}$ is the momentum operator. The constants $c$ and ${{m}_{0}}$ correspond to the speed of light and the rest mass of the electron, respectively. The operator $\hat{V}_{iA}$ describes the nuclear potential, and ${{r}_{ij}}$ denotes the distance between the ${{i}^{\text{th}}}$ and ${{j}^{\text{th}}}$ electrons. \par
The above equation for a many-electron system is solved using the mean-field approximation in the form of the Dirac--Hartree--Fock (DHF) method. The DHF wavefunction $\left| {{\Psi }_{0}}\right\rangle$ is a single determinant made up of 4c spinors, where the large and small components correspond to the electronic and positronic degrees of freedom, respectively. The correlation calculations are usually performed under the no-pair approximation.\cite{sucher1980} \par
The coupled cluster ansatz\cite{cizekCorrelationProblemAtomic1966, CCSD1972, lindroth_1988, visccher1996, Visscher2001} $\left| {{\Psi }_{\text{CC}}} \right\rangle $ based on a DHF wave-function $\left| {{\Psi }_{0}} \right\rangle $ is given as,
\begin{equation}
\label{eqn2}
    \left| {{\Psi }_{\text{CC}}} \right\rangle = {{e}^{{\hat{T}}}}\left| {{\Psi }_{0}} \right\rangle
\end{equation}
The cluster operator $\hat{T}$ in Eq. (\ref{eqn2}) can be expressed as follows,
\begin{equation}
\label{eqn3}
    \hat{T}=\sum\limits_{n}{{{{\hat{T}}}_{n}}}
\end{equation}
${{\hat{T}}}_{n}$ are the set of excitation operators up to $n^{th}$ excitation manifolds and can be written in terms of second quantized operators, 
\begin{equation}
\label{eqn4}
    {{\hat{T}}_{n}}=\sum\limits_{\begin{smallmatrix} 
     i<j<k<... \\ 
     a<b<c<... 
    \end{smallmatrix}}{t_{ijk...}^{abc...}\{\hat{a}_{a}^{\dagger }\hat{a}_{b}^{\dagger }\hat{a}_{c}^{\dagger }...\ {{{\hat{a}}}_{k}}{{{\hat{a}}}_{j}}{{{\hat{a}}}_{i}}...\}}
\end{equation}
Here $t_{ijk...}^{abc...}$ represents the cluster amplitudes, $\hat{a}^{\dagger }$  and $\hat{a}$ are the second quantization creation and annihilation operators, with indices ($i,j,k$ ...) denoting occupied spinors, and ($a,b,c$ ...) representing virtual spinors. 
The cluster amplitudes can be obtained by solving a set of coupled nonlinear equations.
\begin{equation}
\label{eqn5}
    \left\langle  \Psi ^{abc...}_{ijk...} \right|\bar{H}\left| {{\Psi }_{0}} \right\rangle=0
\end{equation}
The relativistic ground-state CC energy can be obtained as
\begin{equation}
\label{eqn6}
    \left\langle  {{\Psi }_{0}} \right|\bar{H}\left| {{\Psi }_{0}} \right\rangle ={{E}_{\text{CC}}}
\end{equation}
Here $\left\langle  \Psi ^{abc...}_{ijk...} \right|$  denote different excited-state determinants. The similarity transformed Hamiltonian $\bar{H}$ is expressed as following,
\begin{equation}
\label{eqn7}
\bar{H} = e^{-\hat{T}} \hat{H}^{\text{4c}} e^{\hat{T}}
\end{equation}
The $\hat{T}$ operators are generally used in singles and doubles approximation (CCSD)
\begin{equation}
\label{eqn8}
  \hat{T}=\hat{T}_{1}+\hat{T}_{2}
\end{equation}
It scales iteratively as $\mathcal{O}(n^6)$ with the basis size. The CCSD method has a storage requirement that scales as $\mathcal{O}(n^4)$ with the basis set, arising from the need to store the two-electron integrals and CC amplitudes.
The full inclusion of triples leads to the CCSDT method
\begin{equation}
\label{eqn9}
  \hat{T}=\hat{T}_{1}+\hat{T}_{2}+\hat{T}_{3}
\end{equation}
It scales iteratively as $\mathcal{O}(n^8)$ with the size of the basis set, while the storage requirement scales as $\mathcal{O}(n^6)$. The primary reason for the increased storage requirement from CCSD to CCSDT is the need to store six-index three-body amplitudes, which can quickly become a computational bottleneck. The CC ground-state formalism can be extended to excited states using the equation-of-motion approach. In this approach, the target-state wavefunction is generated by applying linearized excitation operators to the CC reference state $\left| {{\Psi }_{CC}}\right\rangle$ such that,
\begin{equation}
\label{eqn10}
|\Psi_k\rangle = \hat{R}_k \left|\Psi_{\mathrm{CC}}\right\rangle
\end{equation}
The Hamiltonian acting on this state $|\Psi_k\rangle$ gives its energy,
\begin{equation}
\label{eqn11}
    \hat{H}^{\text{4c}}\left| {{\Psi }_{k}} \right\rangle ={{E}_{k}}\left| {{\Psi }_{k}} \right\rangle 
\end{equation}
For ionized states, the operator $\hat{R}_k$ has the form,
\begin{equation}
\label{eqn12}
    \hat{R}_{k} = \sum_{i} r_{i} \left\{ \hat{a}_{i} \right\}
    + \sum_{\substack{i<j \\ a}} r_{ij}^{a} \left\{ \hat{a}_{a}^{\dagger} \hat{a}_{j} \hat{a}_{i} \right\}+\sum_{\substack{i<j<k \\ a<b}} r_{ijk}^{ab} \left\{ \hat{a}_{a}^{\dagger}\hat{a}_{b}^{\dagger}  \hat{a}_{k}\hat{a}_{j} \hat{a}_{i} \right\}
    + \dots
\end{equation}
The ionization energy can be obtained by solving 
\begin{equation}
\label{eqn13}
    \left[ {{{\bar{H}}}^{\text{}}_{}},{{{\hat{R}}}_{k}} \right]\left| {{\Psi }_{0}} \right\rangle =\left({{E}_{k}}- {{E}_{0}} \right){{\hat{R}}_{k}}\left| {{\Psi }_{0}} \right\rangle ={{\omega }_{k}}{{\hat{R}}_{k}}\left| {{\Psi }_{0}} \right\rangle 
\end{equation}
Here, ${{\omega }_{k}}$ is the ionization energy for the $k^{\text{th}}$ ionized state. Since ${\bar{H}}$ is non-Hermitian, it has both left and right eigenvectors. They are biorthogonally related to each other,
\begin{equation}
\label{eqn14}
  \left\langle  {{\Psi }_{0}} \right|
  \hat{L}_{i} \hat{R}_{j} \left| {{\Psi }_{0}} \right\rangle =\delta_{ij}
 \end{equation}
For ionized states, the $\hat{L}_{k}$ operator is defined as,
\begin{equation}
\label{eqn15}
\hat{L}_k =  \sum_{i} l^{i} 
\left\{ \hat{a}_i^\dagger  \right\}
+ \sum_{\substack{i<j \\ a}} l_{a}^{ij} 
\left\{ \hat{a}_i^\dagger \hat{a}_j^\dagger \hat{a}_a \right\} + \sum_{\substack{i<j<k \\ a<b}} l_{ab}^{ij} 
\left\{ \hat{a}_i^\dagger  \hat{a}_j^\dagger\hat{a}_k^\dagger \hat{a}_b \hat{a}_a\right\} 
+ \dots
\end{equation}
For the calculation of IP, it is sufficient to determine either the right or the left eigenvectors of the similarity-transformed Hamiltonian.  However, the evaluation of transition properties generally requires both sets of eigenvectors.  Equation ~(\ref{eqn13}) is generally solved using Davidson iterative diagonalization technique~\cite{HIRAO1982246}.
Within the IP-EOM-CCSD approximation, Eqs.~\ref{eqn12} and \ref{eqn15} are truncated after the second term. The construction of the so-called ``sigma'' vectors scales as $\mathcal{O}(n^5)$, with a storage requirement comparable to that of the ground-state CCSD method. In the IP-EOM-CCSDT method, Eqs.~\ref{eqn12} and \ref{eqn15} are truncated after the third term. The resulting EOM diagonalization procedure exhibits a computational scaling of $\mathcal{O}(n^7)$, and requires an additional $\mathcal{O}(n^5)$ storage for the three-body EOM amplitudes. The programmable expressions for the relativistic IP-EOM-CCSDT method are provided in the Supporting Information.

\subsubsection*{Approximate Triples}
Most of the computational cost in IP-EOM-CCSDT arises from the construction of the triples amplitudes, which scales as $\mathcal{O}(n^8)$. Numerous iterative and non-iterative approximations to the non-relativistic EOM-CCSDT method have been proposed in the literature.\cite{CCSD_T_EOM1995, CCSD_star_1996, CCSD_star_1997, CC31997, CR_CCSDt2004, CCSDTA_2016} In this work, we employ the approach introduced by Stanton and co-workers\cite{CCSDTA_2016}, in which the ground-state triples amplitudes are approximated at the lowest perturbative order (beyond CCSD).
\begin{equation}
\begin{aligned}
\label{eqn16}
t^{[2]}_3 = \langle \Psi_{ijk}^{abc} |D^{-1}_3 [V, t_2]| \Psi_{0} \rangle
\end{aligned}
\end{equation}
Here ${D}_{3}$ denotes the M{\o}ller--Plesset (MP) denominator in the triples--triples space, i.e., $({D}_{3})_{ijk}^{abc} = \epsilon_{i} + \epsilon_{j} + \epsilon_{k} - \epsilon_{a} - \epsilon_{b} - \epsilon_{c}$, assuming a semi-canonical Fock matrix, where $ f^i_j = \delta_{ij} \epsilon_i, \quad f^a_b = \delta_{ab} \epsilon_a$. The $t^{[2]}_3$ amplitudes are then used to correct the CCSD amplitudes ($t_{1}$ and $t_{2}$) to the lowest perturbative order.
\begin{equation}
\label{eqn17}
\begin{aligned}
t_{1}^{\Delta}
&= \langle \Psi_{i}^{a}|
   D_{1}^{-1}[V,t_{3}^{[2]}]
   |\Psi_{0} \rangle,
\\[4pt]
t_{2}^{\Delta}
&= \langle \Psi_{ij}^{ab}|
   D_{2}^{-1}[F+V,t_{3}^{[2]}]
   |\Psi_{0} \rangle.
\end{aligned}
\end{equation}
Here, $D_{1}=\epsilon_{i}-\epsilon_{a}$ and $D_{2}=\epsilon_{i}+\epsilon_{j}-\epsilon_{a}-\epsilon_{b}$ are the MP denominators for the singles--singles and doubles--doubles spaces, respectively, analogous to $D_{3}$.
The corrected singles and doubles amplitudes can be written as follows.
\begin{equation}
\label{eqn18}
t_{1}^{\mathrm{CCSD(T)(a)}}=t_{1}+ t_{1}^{\Delta}; \quad
t_{2}^{\mathrm{CCSD(T)(a)}}=t_{2}+ t_{2}^{\Delta}
\end{equation}
Using the corrected singles and doubles amplitudes, one can construct the similarity-transformed Hamiltonian $\bar{H}^{0}_{\text{CCSD(T)(a)}}$. This Hamiltonian can be directly used in the EOM calculations. The detailed derivation of the method can be found in the original non-relativistic work of Stanton and co-workers.\cite{CCSDTA_2016} The computational cost of IP-EOM-CCSD(T)(a) consists of four main steps:
(i) the iterative CCSD ground-state calculation, scaling as $\mathcal{O}(n^6)$;
(ii) generation of the perturbative triple amplitudes $t_{3}^{[2]}$;
(iii) construction of $\bar{H}_{\text{CCSD(T)(a)}}$ scaling as $\mathcal{O}(n^7)$; and
(iv) the iterative IP-EOM calculation, also scaling as $\mathcal{O}(n^7)$, which determines the overall cost.
The storage requirement remains the same as that of the full IP-EOM-CCSDT method. To further improve the computational efficiency of IP-EOM calculations, a perturbative approximation, analogous to the ground-state CCSD(T)(a) method, can be extended to the excited-state problem. In this approach, the triples contribution to the excited state is incorporated perturbatively rather than by solving the full excited-state eigenvalue problem. Instead of constructing the full similarity-transformed Hamiltonian $\bar{H}^{0}_{\text{CCSD(T)(a)}}$, the similarity-transformed Hamiltonian $\bar{H}^{\text{corr}}_{\text{CCSD}}$ is constructed using the corrected amplitudes ($T_{1}^{\mathrm{CCSD(T)(a)}}$ and $T_{2}^{\mathrm{CCSD(T)(a)}}$), and the EOM-CCSD equations are subsequently solved with the updated $\bar{H}$ intermediates. The ionization energy is then corrected perturbatively through the so-called ``star'' correction. The contribution of the ground-state triple amplitudes to the 3-internal intermediate ($\bar{H}_{mcij}^{\Delta}$) in the EOM-CCSD equations must also be computed separately and added to the remaining intermediates obtained from the singles and doubles amplitudes. This approach reduces the overall scaling of the EOM part to $\mathcal{O}(n^6)$. The resulting method is denoted IP-EOM-CCSD(T)(a)$^\ast$, with the ``star'' correction\cite{CCSD_star_1996, CCSD_star_1997, CCSDTA_2016} defined as follows,
\begin{equation}
\begin{aligned}
\label{eqn19}
\Delta \omega_{\text{IP}}^{*} = \frac{1}{36} \sum_{i,j,k,a,b,c} 
\frac{
    [\tilde l_{3}]_{\,ab}^{ijk} \cdot [\tilde r_{3}]_{\,ijk}^{ab}
}{
    \epsilon_i + \epsilon_j + \epsilon_k - \epsilon_a - \epsilon_b - \epsilon_c 
    + \omega_{\text{IP}}^{\text{EOM-CCSD}}[\bar{H}_{(T)(a)}]
},
\end{aligned}
\end{equation}
in which,
\begin{equation}
\begin{aligned}
\label{eqn20}
[\tilde l_{3}]_{\,ab}^{ijk} &=
\left\langle \Psi_{0} \middle| (L_{1}+L_{2})\bar{H}^{[1]} \middle| \Psi_{ijk}^{ab} \right\rangle, \\
[\tilde r_{3}]_{\,ijk}^{ab} & = \left\langle \Psi_{ijk}^{ab}  \middle| (\bar{H}^{[2]}R_{1}+\bar{H}^{[1]}R_{2}) \middle| \Psi_{0} \right\rangle
\end{aligned}
\end{equation}
and the $\epsilon$'s denote molecular spinor energies. $\bar{H}^{[1]}$ and $\bar{H}^{[2]}$ denote the similarity-transformed Hamiltonian corrected to first and second order in M{\o}ller--Plesset perturbation theory, respectively. The explicit programmable expressions for IP-EOM-CCSD(T)(a)$^\ast$
are given in the Appendix. An additional advantage of the relativistic IP-EOM-CCSD(T)(a)$^\ast$ approach is that it does not require storage of three-body amplitudes for either the ground or the ionized state, so its storage requirements are comparable to those of the relativistic IP-EOM-CCSD method.
One can obtain a further approximation to IP-EOM-CCSD(T)(a)$^\ast$ by neglecting the effect of ground-state triples in the calculations. This leads to the original IP-EOM-CCSD$^\ast$ method of Stanton and co-workers.\cite{CCSD_star_1996, CCSD_star_1997}

\begin{comment}
 Rather than solving the full IP-EOM-CCSDT(a) eigenvalue problem, the excitation space is partitioned into $P$ and $Q$ subspaces, where $P$ includes single and double excitations, and $Q$ encompasses triples and higher. The $P$ space is treated iteratively via standard EOM-CCSD using the similarity-transformed Hamiltonian $\bar{H}^{0}_{\text{CCSD(T)(a)}}$. The contribution from the $Q$ space is incorporated perturbatively as a non-iterative correction—termed the “star” correction—mirroring the approach used in ground-state CCSD(T)(a).   
\end{comment}

\subsection{X2CAMF Approximation}
Further computational simplification can be achieved by transforming to a 2c representation. The DC Hamiltonian, expressed in second-quantized form, can be written as

\begin{equation}
\begin{aligned}
\label{eqn21}
    \hat{H}^{\text{4c}} = \sum_{pq}{h^{\text{4c}}_{pq}\hat{a}_p^{\dagger}\hat{a}_q}
    + \frac{1}{4}\sum_{pqrs}{g^{\text{4c,SF}}_{pqrs}\hat{a}_p^{\dagger}\hat{a}_q^{\dagger}\hat{a}_s\hat{a}_r}
    + \frac{1}{4}\sum_{pqrs}{g^{\text{4c,SD}}_{pqrs}\hat{a}_p^{\dagger}\hat{a}_q^{\dagger}\hat{a}_s\hat{a}_r}.
\end{aligned}
\end{equation}
 The two-electron part is decomposed into spin-free (SF) and spin-dependent (SD) contributions.\cite{spin_sep_1994} The spin-dependent component is treated within the atomic mean-field (AMF) approximation, exploiting the localized nature of the spin--orbit interactions.
\begin{equation}
\label{eqn22}
    \frac{1}{4}\sum_{pqrs}{g^{\text{4c,SD}}_{pqrs}\hat{a}_p^{\dagger}\hat{a}_q^{\dagger}\hat{a}_s\hat{a}_r}
    \approx \sum_{pq}{g^{\text{4c,AMF}}_{pq}\hat{a}_p^{\dagger}\hat{a}_q}=\sum_{pqiA}{n_{iA}}g^{\text{4c,SD}}_{p_{iA}q_{iA}}\hat{a}_p^{\dagger}\hat{a}_q
\end{equation}
where $A$ labels the distinct atoms in the molecule, $i$ denotes occupied spinors localized on atom $A$, and $n_{iA}$ denotes their corresponding occupation numbers. Substituting Eq. (\ref{eqn22}) in Eq. (\ref{eqn21}), we get,
\begin{equation}
\label{eqn23}
    %\begin{align}
    \hat{H}^{\text{4c}} = \sum_{pq}{h^{\text{4c}}_{pq}\hat{a}_p^{\dagger}\hat{a}_q}
    + \frac{1}{4}\sum_{pqrs}{g^{\text{4c,SF}}_{pqrs}\hat{a}_p^{\dagger}\hat{a}_q^{\dagger}\hat{a}_s\hat{a}_r}
    + \sum_{pq}{g^{\text{4c,AMF}}_{pq}\hat{a}_p^{\dagger}\hat{a}_q}
    %\end{align}
\end{equation}
The 4c Hamiltonian can be transformed to a 2c representation using the relationship between the large and small component coefficients through the $X$ matrix, and between the large component and the 2c wave function through the $R$ matrix.
\begin{equation}
\label{eqn24}
    C^S=XC^L
\end{equation}
\begin{equation}
\label{eqn25}
    C^L=RC^{\text{2c}}
\end{equation}
The spin-free contribution $g^{\text{4c,SF}}$ reduces to the non-relativistic two-electron integrals $g^{\text{NR}}$ when scalar two-electron picture-change (2e-pc) contributions are neglected.
%\begin{equation}
%\label{eqn23}
%    g^{\text{4c,SF}}_{pqrs} \approx g^{\text{NR}}_{pqrs}
%\end{equation}
The 4c Hamiltonian, after transformation, becomes the 2c X2CAMF Hamiltonian\cite{Daoling2009}
\begin{equation}
\label{eqn26}
%\begin{align}
    \hat{H}^{\text{X2CAMF}} = \sum_{pq}{h^{\text{X2C}}_{pq}\hat{a}_p^{\dagger}\hat{a}_q}
    + \frac{1}{4}\sum_{pqrs}{g^{\text{NR}}_{pqrs}\hat{a}_p^{\dagger}\hat{a}_q^{\dagger}\hat{a}_s\hat{a}_r}
    + \sum_{pq}{g^{\text{2c,AMF}}_{pq}\hat{a}_p^{\dagger}\hat{a}_q}
%\end{align}
\end{equation}
The above Hamiltonian can be written in terms of an effective one-electron operator and a non-relativistic two-electron operator as:
\begin{equation}
\label{eqn27}
    \hat{H}^{\text{X2CAMF}} = \sum_{pq}{h^{\text{X2CAMF}}_{pq}\hat{a}_p^{\dagger}\hat{a}_q}
    + \frac{1}{4}\sum_{pqrs}{g^{\text{NR}}_{pqrs}\hat{a}_p^{\dagger}\hat{a}_q^{\dagger}\hat{a}_s\hat{a}_r}
\end{equation}
where $h^{\text{X2CAMF}} = h^{\text{X2C}}+g^{\text{2c,AMF}}$ is the effective one-electron operator. The main advantage of this Hamiltonian is that it completely avoids the explicit construction of relativistic two-electron integrals.\cite{AMF2018, AMF2022, AMF_2022_2} In the current work, the two-electron integrals are evaluated using the Cholesky decomposition technique.\cite{chol2024, chol_rel2024}
%In X2CAMF-based CC methods, the similarity-transformed Hamiltonian %can be written as
%\begin{equation}
%\label{eqn26}
%    \bar{H}=e^{-\hat{T}}\hat{H}^{\text{X2CAMF}}e^{\hat{T}}
%\end{equation}

\subsection{Cholesky Decomposition}
Within the Cholesky decomposition framework\cite{CD_orginal}, electron-repulsion integrals (ERIs) are approximated as
\begin{equation}
\label{eqn28}
\left(\mu\nu|\kappa\lambda\right) \approx \sum_{P=1}^{n_{\text{CD}}} L_{\mu\nu}^{P} L_{\kappa\lambda}^{P},
\end{equation}
where $\mu, \nu, \kappa, \lambda$ denote atomic spinor indices, $L_{\mu\nu}^{P}$ are the Cholesky vectors, and $n_{\text{CD}}$ is their rank.
In the present implementation,  an iterative CD algorithm is employed\cite{CD_CCSD_2025}. At each stage, the largest diagonal element of the ERI matrix is identified, and the procedure is repeated until this element falls below a user-defined threshold ($\tau$). Once constructed, the Cholesky vectors are transformed into the molecular spinor basis as
\begin{equation}
\label{eqn29}
L_{pq}^{P} = \sum_{\mu\nu} C_{\mu p}^{*} L_{\mu\nu}^{P} C_{\nu q}.
\end{equation}
The molecular-orbital Cholesky vectors provide a compact route to evaluate anti-symmetrized two-electron integrals:
\begin{equation}
\label{eqn30}
\left\langle pq||rs\right\rangle = \sum_{P=1}^{n_{\text{CD}}} \big( L_{pr}^{P}L_{qs}^{P} - L_{ps}^{P}L_{qr}^{P} \big).
\end{equation}
In the present implementation, integrals with three or four virtual indices (e.g., $\langle ab||ci \rangle$ and $\langle ab||cd \rangle$) are generated on the fly rather than stored explicitly, to reduce memory/storage requirements. All other integrals are constructed once and stored. 
\subsection{Natural Spinors} 
Frozen natural spinors (FNS) have emerged as an efficient framework for reducing the computational cost of relativistic wavefunction-based calculations.\cite{fns_ccsd_2022, Visccher_2022, chakraborty2025low, sujan} Among the multiple variants of FNS,\cite{fns_ccsd_2022, tamoghnass, fnsplus} the MP2-based natural spinors provide an optimal balance between accuracy and computational cost for describing singly ionized states within the relativistic EOM-CC framework.\cite{fns_ip_2022} The construction of MP2-based frozen natural spinors proceeds as follows:
\begin{enumerate}
  \item A one-particle reduced density matrix (1-RDM) is constructed using the X2CAMF-MP2 method, and its virtual–virtual block $\Gamma_{ab}$ is extracted. 
  \begin{equation}
    \label{eqn31}
    \Gamma_{ab}=\frac{1}{2}\sum_{cij} {\frac{\langle ac||ij \rangle \langle ij||bc \rangle}{ (\epsilon_i + \epsilon_j - \epsilon_a - \epsilon_c)(\epsilon_i + \epsilon_j - \epsilon_b - \epsilon_c)}}
    \end{equation}
  
  \item The matrix $\Gamma_{ab}$ is diagonalized to obtain the eigenvectors   $V$ (virtual natural spinors) and the corresponding eigenvalues $\eta$ (occupation numbers).
      \begin{equation}
    \label{eqn32}
        \Gamma_{ab}V=V\eta
    \end{equation}

  \item The virtual–virtual block of the Fock matrix  $F_{vv}$ is transformed to the truncated natural spinor basis $F^{\mathrm{NS}}_{vv}$ by applying a predetermined threshold (cutoff) to the occupation numbers.
  \begin{equation}
    \label{eqn33}
        F^{\mathrm{NS}}_{vv}={{\tilde{V}}^{\dagger }}F_{vv}\tilde{V}
    \end{equation}

  \item The matrix $F^{\mathrm{NS}}_{vv}$ is diagonalized to obtain the semicanonical virtual natural spinors $\tilde{Z}$ and their associated orbital energies $\tilde{\epsilon}$.
  \begin{equation}
    \label{eqn34}
        F^{\mathrm{NS}}_{vv}\tilde{Z}=\tilde{Z}\tilde{\epsilon }
    \end{equation}
    
  \item The full canonical space is transformed to the semi-canonical space using the transformation matrix $U$.
      \begin{equation}
    \label{eqn35}
        U=\tilde{V}\tilde{Z}
    \end{equation}
\end{enumerate}

Within the compact virtual space defined by the FNS scheme, both the number of floating-point operations and the memory requirements for subsequent correlation calculations are significantly reduced.

\subsection{Implementation and Computational Details}
The methods developed in this work are denoted as FNS-IP-EOM-CCSD*, FNS-IP-EOM-CCSD(T)(a)*, and FNS-IP-EOM-CCSD(T)(a). These methods have been implemented in the development version of our in-house quantum chemistry software, BAGH\cite{duttaBAGHQuantumChemistry2025_a}. The package is primarily written in Python, with performance-critical components optimized in Cython and Fortran. BAGH provides interfaces to PySCF\cite{pyscf1, pyscf2, pyscf3}, GAMESS-US\cite{gamess}, socutils\cite{wangXubwaSocutils2025_a}, and DIRAC\cite{DIRAC2025}. The X2CAMF-HF calculations are carried out using the socutils package through its integration with BAGH. Additionally, the FNS-IP-EOM-CCSDT method has been implemented to facilitate direct comparison with the aforementioned methods. 

\begin{figure}[htbp]
    \centering
    \includegraphics[width=0.45\textwidth]{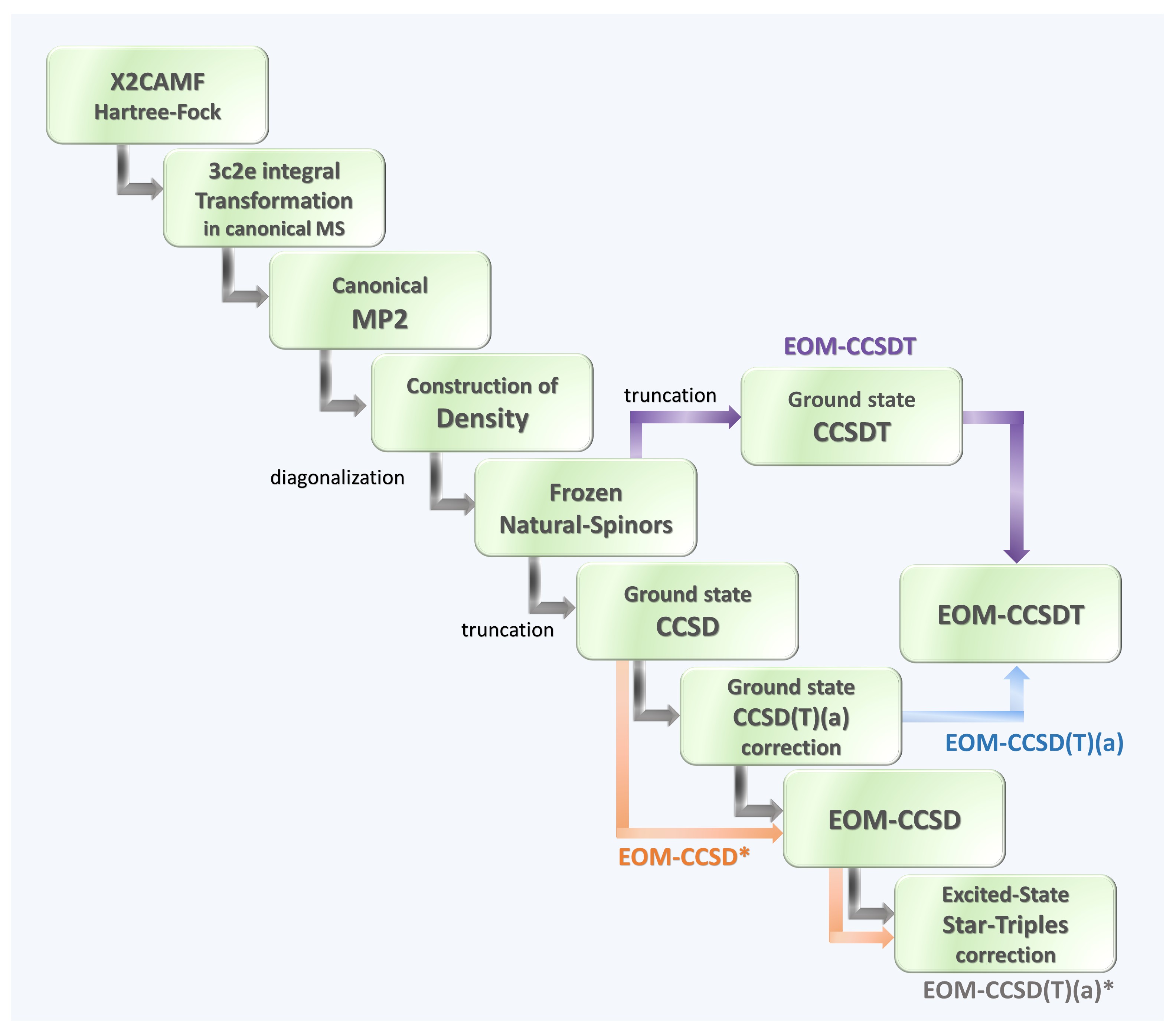}
    \caption{\justifying{Schematic representation for different CD-based X2CAMF-FNS methods.}}
    \label{fig:flow_chart}
\end{figure}

Fig.~\ref{fig:flow_chart} provides a schematic representation of the complete computational workflow, detailing the construction of the FNS space and the subsequent evaluation of electron correlation effects. The figure presents a sequential overview of the algorithmic procedure employed by all methods, with corresponding color coding. The IP-EOM-CCSD(T)(a)$^\ast$ and IP-EOM-CCSD(T)(a) approaches employ identical ground-state CC treatments and differ only in the form of the similarity-transformed Hamiltonian ($\bar{H}$) and subsequent EOM calculations. In contrast, IP-EOM-CCSD(T)(a) and IP-EOM-CCSDT share the same $\bar{H}$ construction and EOM framework. Finally, EOM-CCSD$^\ast$ and EOM-CCSD(T)(a)$^\ast$ utilize identical $\bar{H}$ and EOM operators but differ in the underlying ground-state correlation treatment. All calculations in this study were performed using the CD-X2CAMF Hamiltonian, unless stated otherwise. The frozen-core approximation was employed throughout.
\section{\label{sec:level4}Results and Discussion}

% FIGURE: flowchart
\subsection{Choice of Truncation Threshold}
\begin{figure}[htbp]
    \centering
    \includegraphics[width=0.45\textwidth]{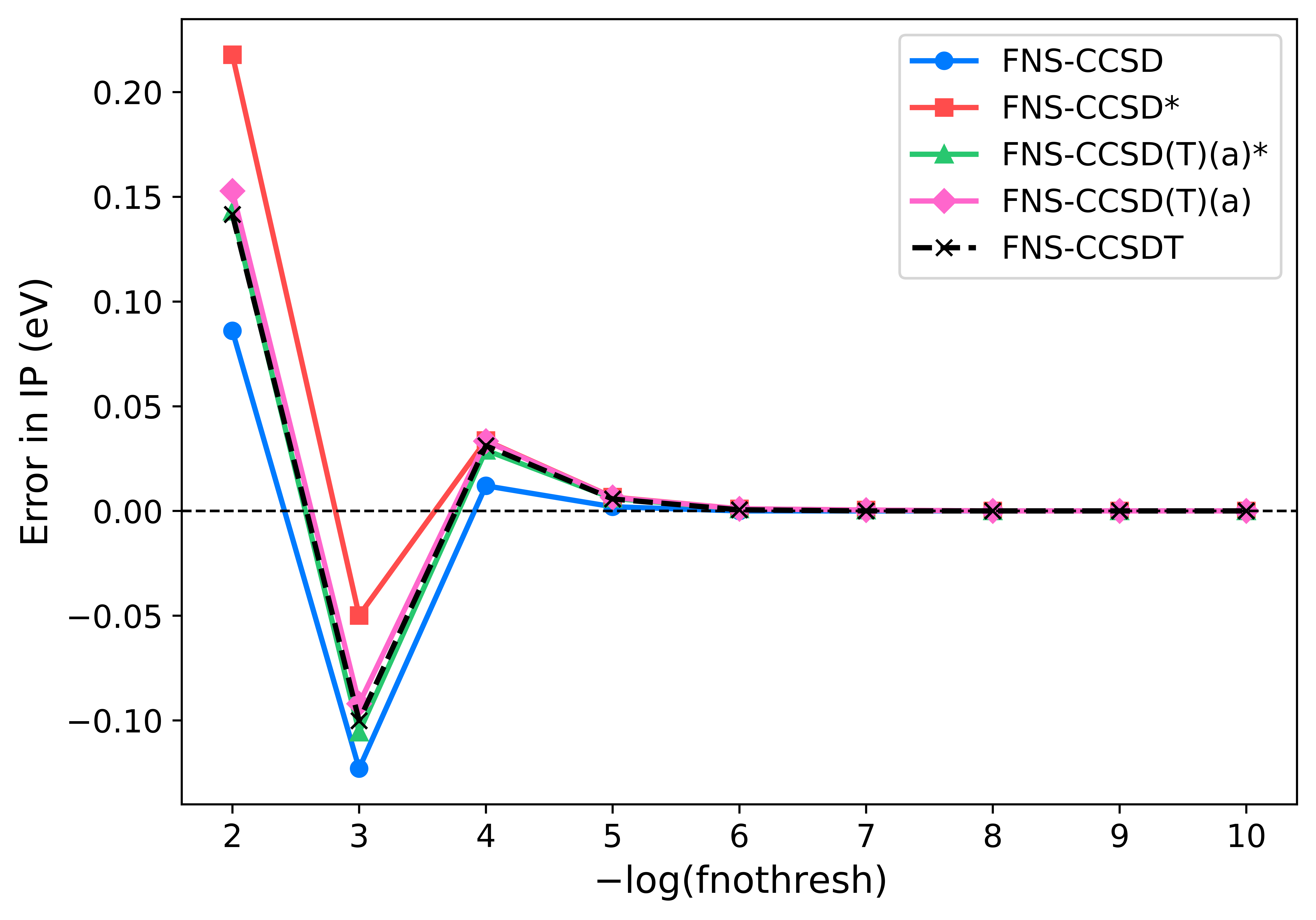}
    \caption{\justifying{Convergence of the ionization energy calculated using different EOM-CC methods with respect to the FNS truncation threshold ($10^{−n}$) for the HCl molecule.}}
    \label{fig:error_thresh}
\end{figure}

The accuracy of FNS-based IP-EOM-CC methods depends on the FNS threshold used in the calculations,\cite{fns_ccsd_2022, fns_ip_2022} as well as on the CD threshold when the CD-X2CAMF approximation is employed. Our previous studies\cite{CD_CCSD_2025, CD_IP_2025} have shown that the computed IP values are not very sensitive to the CD threshold (above a certain limit) and that a value of $10^{-3}$ provides an optimal balance between computational cost and accuracy. Therefore, we have kept the CD threshold fixed at $10^{-3}$ throughout the present study. For the FNS threshold, we have performed benchmark calculations on the HCl molecule using the dyall.v3z basis set. Fig.~\ref{fig:error_thresh} shows the deviation (with respect to results obtained with the untruncated canonical virtual space) as a function of the FNS truncation threshold for different triples correction schemes within the relativistic IP-EOM-CC framework. All methods exhibit the same convergence behavior, and the IP of HCl is converged at a threshold of $10^{-5}$. Accordingly, this FNS threshold value was adopted for all subsequent calculations.
% FIGURE: Threshold

\subsection{The comparison of accuracy in different partial triples correction schemes}
\begin{figure}[htbp]
    \centering
    \includegraphics[width=0.45\textwidth]{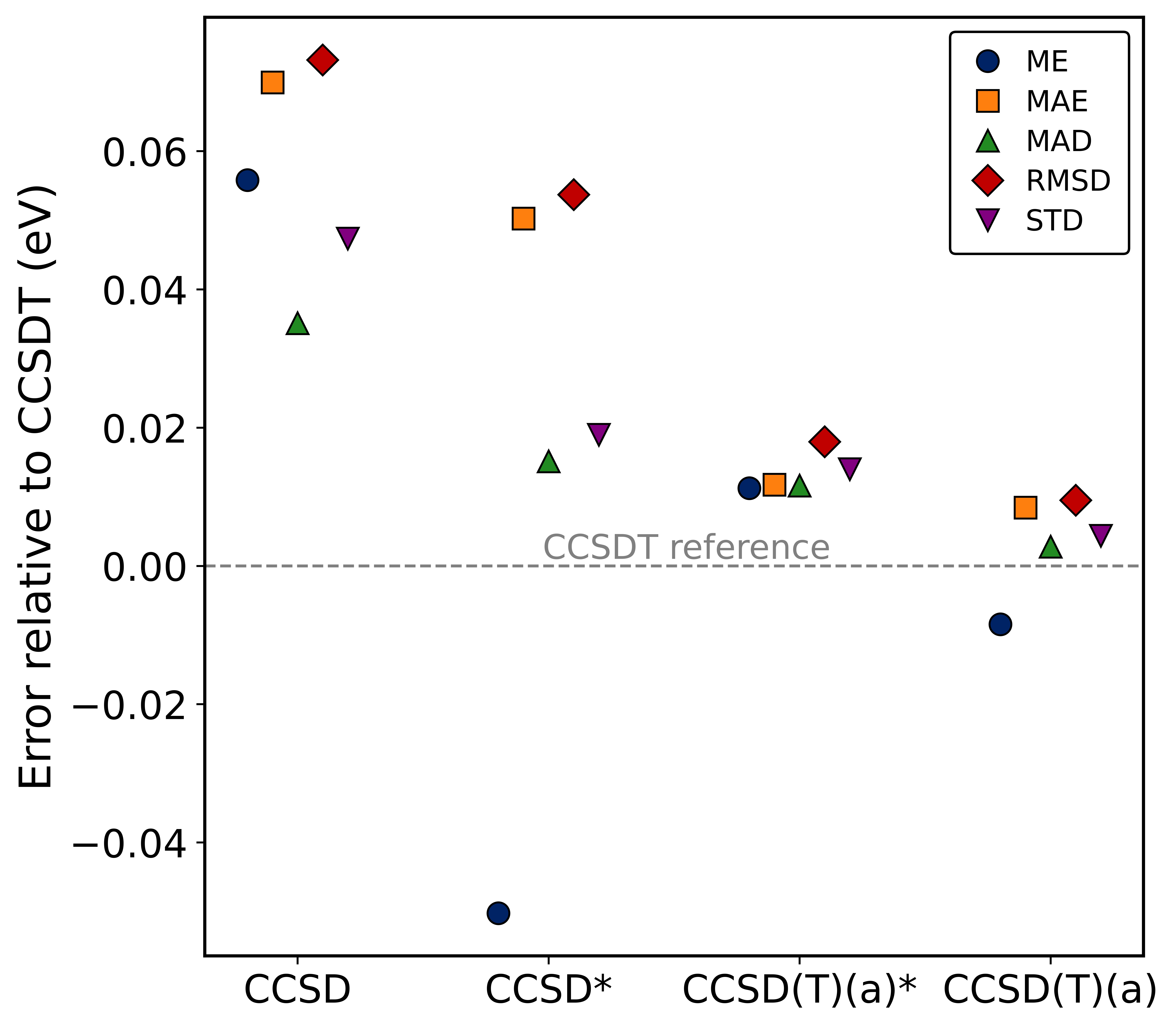}
    \caption{\justifying{The deviations in ionization energies of atoms and anions obtained from different IP-EOM-CC methods relative to the reference IP-EOM-CCSDT values. The dyall.v3z basis set was employed for all atoms.}}
    \label{fig:error_CCSDT}
\end{figure}

To evaluate the accuracy of the methods discussed above, vertical ionization energies of $^{1}S_{0}$ atoms, including halide ions ($X^{-}$, $X$ = F, Cl, Br, and I) and noble gases (Ne, Ar, Kr, and Xe), were benchmarked against FNS-IP-EOM-CCSDT calculations performed with the dyall.v3z basis set. The deviations in the energies of the $^{2}P_{3/2}$ and $^{2}P_{1/2}$ states relative to the reference EOM-CCSDT IP values were systematically analyzed, and the corresponding statistical measures are summarized in Tables~S1 and S2 and plotted in Figure~\ref{fig:error_CCSDT}. Figure~\ref{fig:error_CCSDT} shows that EOM-CCSD exhibits a pronounced positive mean error (ME) of approximately 0.06~eV, indicating systematic overestimation of the IP. In contrast, EOM-CCSD$^\ast$ yields a negative ME of about $-0.05$~eV, reflecting systematic underestimation. Both EOM-CCSD(T)(a)$^\ast$ and EOM-CCSD(T)(a) produce substantially smaller ME values of approximately 0.01~eV and $-0.01$~eV, respectively, demonstrating much closer agreement with the EOM-CCSDT reference. Notably, methods employing asymmetric treatments of the ground and excited states, such as EOM-CCSD$^\ast$ and EOM-CCSD(T)(a), tend to underestimate IP, whereas methods with a more balanced treatment, including EOM-CCSD and EOM-CCSD(T)(a)$^\ast$, tend to overestimate them. Similar trends are observed in the mean absolute error (MAE). EOM-CCSD shows the largest MAE (approximately 0.07~eV), followed by EOM-CCSD$^\ast$ (about 0.05~eV). A significant improvement is obtained with EOM-CCSD(T)(a)$^\ast$, for which the MAE decreases to roughly 0.01~eV, while EOM-CCSD(T)(a) yields a comparable value, indicating consistently high accuracy across the dataset. The reduction in error is further reflected in the root-mean-square deviation (RMSD) and standard deviation (STD). EOM-CCSD exhibits the largest dispersion of the errors, with an RMSD of approximately 0.07~eV and an STD of about 0.05~eV. For EOM-CCSD$^\ast$, the RMSD is reduced to around 0.05~eV, together with a corresponding decrease in the STD, indicating improved consistency. A further reduction is observed for EOM-CCSD(T)(a)$^\ast$, where both RMSD and STD fall in the range of 0.01--0.02~eV. The smallest dispersion is obtained with EOM-CCSD(T)(a), with an RMSD close to 0.01~eV and an STD of approximately 0.004~eV, reflecting tightly clustered errors and minimal residual bias. The maximum absolute deviation (MAD) follows the same overall trend. EOM-CCSD exhibits the largest MAD (about 0.04~eV), which decreases to approximately 0.02~eV for EOM-CCSD$^\ast$. For EOM-CCSD(T)(a)$^\ast$, the MAD remains well below 0.02~eV and is close to 0.01~eV, while the smallest MAD (about 0.003~eV) is obtained with EOM-CCSD(T)(a). Overall, this analysis demonstrates that IP-EOM-CCSD(T)(a)$^\ast$ and IP-EOM-CCSD(T)(a) provide efficient and reliable approximations to the full IP-EOM-CCSDT treatment, achieving substantially reduced systematic and statistical errors across the benchmark set.
% FIGURE: Errors (CCSDT)
\subsection{Comparison with Experimental IP}
For further validation, we have calculated the vertical valence IPs of hydrogen halides (HX; X= F, Cl, Br, I and At) using the partial triples correction schemes developed in this work. The valence dyall basis sets (dyall.v$n$z; $n = 2, 3, 4$) were employed, and the results were extrapolated to the complete basis set (CBS) limit using the Peterson--Dunning scheme.\cite{dunning1994} From Table~\ref{table:1}, it is evident that both IP-EOM-CCSD(T)(a) and IP-EOM-CCSD(T)(a)$^\ast$ show good agreement with experimental results in most cases, with CCSD(T)(a) providing the closest overall match at the CBS level. In general, all of these methods slightly overestimate the IPs relative to experiment, whereas CCSD$^\ast$ underestimates them in some cases. This trend is broadly consistent with that observed for atoms, where the calculated IPs were compared with full IP-EOM-CCSDT results. Although IP-EOM-CCSD(T)(a) appears to be the most accurate method, it is also the most computationally demanding among those considered. In contrast, IP-EOM-CCSD(T)(a)$^\ast$ reproduces comparable results at significantly reduced computational cost. Thus, IP-EOM-CCSD(T)(a)$^\ast$ provides an optimal balance between cost and accuracy, and has been employed for all subsequent calculations in this manuscript.
\begin{table*}[ht!]
\caption{\justifying{CBS ionization potentials (eV) of hydrogen halides computed using different IP-EOM-CC methods and compared with experimental values.}}
\begin{ruledtabular}
%%\begin{tabular}{l l d{3} d{3} d{3} d{3} d{3}}
\begin{tabular}{l l D{.}{.}{3} D{.}{.}{3} D{.}{.}{3} D{.}{.}{3} D{.}{.}{3}}
Molecule & Ionic State & \multicolumn{1}{c}{CCSD} & \multicolumn{1}{c}{CCSD$^\ast$} & \multicolumn{1}{c}{CCSD(T)(a)$^\ast$} & \multicolumn{1}{c}{CCSD(T)(a)} & \multicolumn{1}{c}{Experiment} \\
\hline
\multirow{3}{*}{HF} 
 & $^2\Pi_{1/2}$   & 16.134 & 16.134 & 16.217 & 16.167 & 16.120\cite{Banna1975} \\
 & $^2\Pi_{3/2}$   & 16.175 & 16.173 & 16.257 & 16.206 & \multicolumn{1}{c}{---} \\
 & $^2\Sigma_{1/2}$& 20.063 & 20.055 & 20.138 & 20.089 & 19.890\cite{Banna1975} \\
\multirow{3}{*}{HCl} 
 & $^2\Pi_{1/2}$   & 12.819 & 12.691 & 12.788 & 12.772 & 12.744\cite{HCL1998} \\
 & $^2\Pi_{3/2}$   & 12.902 & 12.773 & 12.870 & 12.854 & 12.830\cite{HCL1998} \\
 & $^2\Sigma_{1/2}$& 16.831 & 16.696 & 16.791 & 16.763 & \multicolumn{1}{c}{---} \\
\multirow{3}{*}{HBr} 
 & $^2\Pi_{1/2}$   & 11.739 & 11.614 & 11.702 & 11.694 & 11.680\cite{HBr1992} \\
 & $^2\Pi_{3/2}$   & 12.069 & 11.943 & 12.030 & 12.021 & 11.980\cite{HBr1992} \\
 & $^2\Sigma_{1/2}$& 15.837 & 15.706 & 15.797 & 15.776 & 15.650\cite{HBr1992} \\
\multirow{3}{*}{HI} 
 & $^2\Pi_{1/2}$   & 10.440 & 10.314 & 10.401 & 10.397 & 10.388\cite{HI_1975} \\
 & $^2\Pi_{3/2}$   & 11.096 & 10.972 & 11.058 & 11.051 & 11.047\cite{HI_1975} \\
 & $^2\Sigma_{1/2}$& 14.470 & 14.336 & 14.430 & 14.405 & \multicolumn{1}{c}{---} \\
\multirow{3}{*}{HAt} 
 & $^2\Pi_{1/2}$   & 9.332  & 9.150  & 9.263  & 9.296  & \multicolumn{1}{c}{---} \\
 & $^2\Pi_{3/2}$   & 11.061 & 10.876 & 10.993 & 11.014 & \multicolumn{1}{c}{---} \\
 & $^2\Sigma_{1/2}$& 14.254 & 14.053 & 14.177 & 14.138 & \multicolumn{1}{c}{---} \\
\end{tabular}
\end{ruledtabular}
\label{table:1}
\end{table*}

\subsubsection*{Comparison with 4-component results}
\begin{figure}[htbp]
    \centering
    \includegraphics[width=0.45\textwidth]{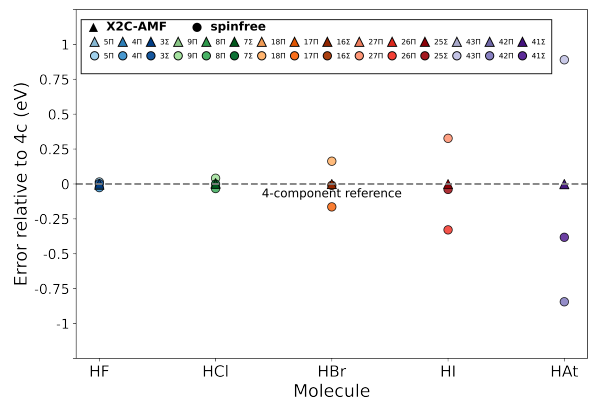}
    \caption{\justifying{Deviations in ionization energies computed using the X2CAMF and spin-free X2C Hamiltonians relative to the four-component DC reference, for FNS-IP-EOM-CCSD(T)(a)$^\ast$. The dyall.v4z basis set was used for all systems.}}
    \label{fig:error_x2c_amf_spinfree}
\end{figure}

The accuracy of the X2CAMF framework for ionization energy calculations at the IP-EOM-CCSD level has been conclusively established in an earlier work from our group.\cite{CD_IP_2025} The present results confirm that this accuracy is preserved upon inclusion of perturbative triple-excitation corrections. Using the FNS-IP-EOM-CCSD(T)(a)$^\ast$ method, we have evaluated the IPs of hydrogen halides employing three relativistic Hamiltonians: the 4c DC Hamiltonian, the spin-free X2C Hamiltonian, and the X2CAMF Hamiltonian. All calculations were performed with the dyall.v4z basis set. The resulting deviations are analyzed through scatter plots. As illustrated in Fig.~\ref{fig:error_x2c_amf_spinfree}, the X2CAMF Hamiltonian reproduces the 4c valence IPs with negligible deviations, indicating an accurate treatment of both scalar relativistic and spin--orbit effects. In contrast, the spin-free X2C Hamiltonian leads to substantially larger deviations, highlighting the importance of explicit spin--orbit coupling for reliable ionization energy predictions.
% FIGURE: Spin-free vs x2c-mf
The deviation in IPs obtained with the spin-free X2C Hamiltonian is small for lighter systems such as HF and HCl but increases significantly for heavier halogens. This behavior reflects the expected growth of relativistic contributions with increasing nuclear charge, as both scalar relativistic and spin-dependent effects scale with atomic number. In heavier elements, spin--orbit coupling plays an increasingly important role in determining the electronic structure, and its omission in the spin-free X2C formulation leads to noticeable errors in the computed ionization energies. In contrast, the X2CAMF framework incorporates both scalar relativistic and spin--orbit effects, yielding IPs in excellent agreement with the 4c reference values across the series. These results confirm that the X2CAMF approach remains a reliable alternative to the 4c Hamiltonian for ionization energy calculations beyond the singles--doubles level. A notable observation is that, for most systems, the first and second IPs (of $\Pi$ symmetry) show comparatively larger deviations when computed with the spin-free Hamiltonian, while the third IP (of $\Sigma$ symmetry) exhibits substantially smaller deviations, since it derives from a molecular orbital with negligible spin--orbit splitting.

\subsection{Ionization of Di-Halogen Molecules}
To further assess the performance of FNS-IP-EOM-CCSD(T)(a)$^\ast$, we computed the vertical valence IPs of dihalogen molecules (X$_2$; X = F, Cl, Br, I) and compared the results with those of CCSD and CCSD$^\ast$, as well as with available experimental data. For F$_2$ and Cl$_2$, uncontracted core-valence Dunning basis sets (aug-cc-pCVNZ; $N = $ D, T, Q) were employed, whereas for Br$_2$ and I$_2$, core-valence dyall basis sets (dyall.v$n$z; $n = 2, 3, 4$) were used. The IPs were extrapolated to the complete basis set (CBS) limit using the Peterson--Dunning scheme.\cite{dunning1994} As shown in Table~\ref{tab:CBS_Di_Hal}, all methods generally overestimate the IPs, with only a few exceptions. Among them, CCSD(T)(a)$^\ast$ provides the most accurate description of the first valence IPs for nearly all dihalogen molecules considered, followed by CCSD and CCSD$^\ast$. In almost all cases, the first IPs obtained with CCSD(T)(a)$^\ast$ lie within the reported experimental uncertainties. For the third IPs, however, CCSD$^\ast$ shows the closest agreement with experiment, followed by CCSD(T)(a)$^\ast$ and CCSD. None of the methods reproduces the third IPs within the experimental error bars. The slightly improved performance of CCSD$^\ast$ relative to CCSD(T)(a)$^\ast$ for the third IPs can be attributed to fortuitous error cancellation.

\begin{table}[h!]
\centering
\caption{CBS ionization potentials (eV) of dihalogen molecules computed using different CC methods, compared with experiment.}
\label{tab:CBS_Di_Hal}
\begin{tabular}{llcccc}
\toprule
Molecule & Ionic state & CCSD & CCSD$^\ast$ & CCSD(T)(a)$^\ast$ & Experiment\footnotemark[1] \\
\midrule
F$_2$  
& ${}^{2}\Pi_{u,3/2}$ & 15.711 & 15.623 & 15.773 & 15.83(0.01) \\
& ${}^{2}\Pi_{u,1/2}$ & 15.756 & 15.668 & 15.818 &             \\
& ${}^{2}\Pi_{g,3/2}$ & 19.051 & 18.869 & 19.008 & 18.80(0.02) \\
\midrule
Cl$_2$ 
& ${}^{2}\Pi_{u,3/2}$ & 11.624 & 11.451 & 11.582 & 11.59(0.01) \\
& ${}^{2}\Pi_{u,1/2}$ & 11.715 & 11.541 & 11.673 &             \\
& ${}^{2}\Pi_{g,3/2}$ & 14.575 & 14.349 & 14.467 & 14.40(0.02) \\
\midrule
Br$_2$ 
& ${}^{2}\Pi_{u,3/2}$ & 10.690 & 10.452 & 10.596 & 10.56(0.01) \\
& ${}^{2}\Pi_{u,1/2}$ & 11.041 & 10.799 & 10.945 &             \\
& ${}^{2}\Pi_{g,3/2}$ & 13.016 & 12.718 & 12.856 & 12.77(0.02) \\
\midrule
I$_2$  
& ${}^{2}\Pi_{u,3/2}$ & 9.483  & 9.165  & 9.351  & 9.35(0.01)  \\
& ${}^{2}\Pi_{u,1/2}$ & 10.120 & 9.789  & 9.983  &             \\
& ${}^{2}\Pi_{g,3/2}$ & 11.362 & 10.985 & 11.167 & 11.01(0.02) \\
\midrule
ME    &              & 0.150  & -0.087 & 0.061 \\
MAE   &              & 0.180  & 0.105    & 0.078 \\
MAD   &              & 0.154  & 0.080    & 0.072 \\
RMSD  &              & 0.213  & 0.122    & 0.103 \\
STD   &              & 0.152  & 0.086    & 0.083 \\
\bottomrule
\end{tabular}
\footnotetext[1]{The experimental values are taken from Ref.~\cite{Di_hal_IP_1971}.}
\end{table}

The statistical analysis summarized at the bottom of Table~\ref{tab:CBS_Di_Hal} is based on the experimental first and third IPs. All statistical indicators consistently demonstrate that CCSD(T)(a)$^\ast$ provides the most accurate overall description, yielding the smallest ME, MAE, MAD, RMSD, and STD. The near-zero ME indicates the absence of any significant systematic bias, while the low MAE and MAD values reflect uniformly improved agreement with experimental data across the dihalogen series. In contrast, CCSD exhibits the largest deviations in all statistical measures, reflecting a systematic overestimation of the IPs. The CCSD$^\ast$ approach represents a substantial improvement over CCSD and achieves error magnitudes comparable to, albeit slightly larger than, those obtained with CCSD(T)(a)$^\ast$. Overall, these results underscore the crucial role of triple-excitation effects in achieving quantitatively reliable IPs for dihalogen molecules. Notably, the significant improvement already observed at the CCSD$^\ast$ level suggests that perturbative corrections applied at the excited-state level effectively capture the dominant triple-excitation contributions to the IPs of dihalogens.

\subsection{Computational Timing}
\begin{figure}[htbp]
    \centering
    \includegraphics[width=0.45\textwidth]{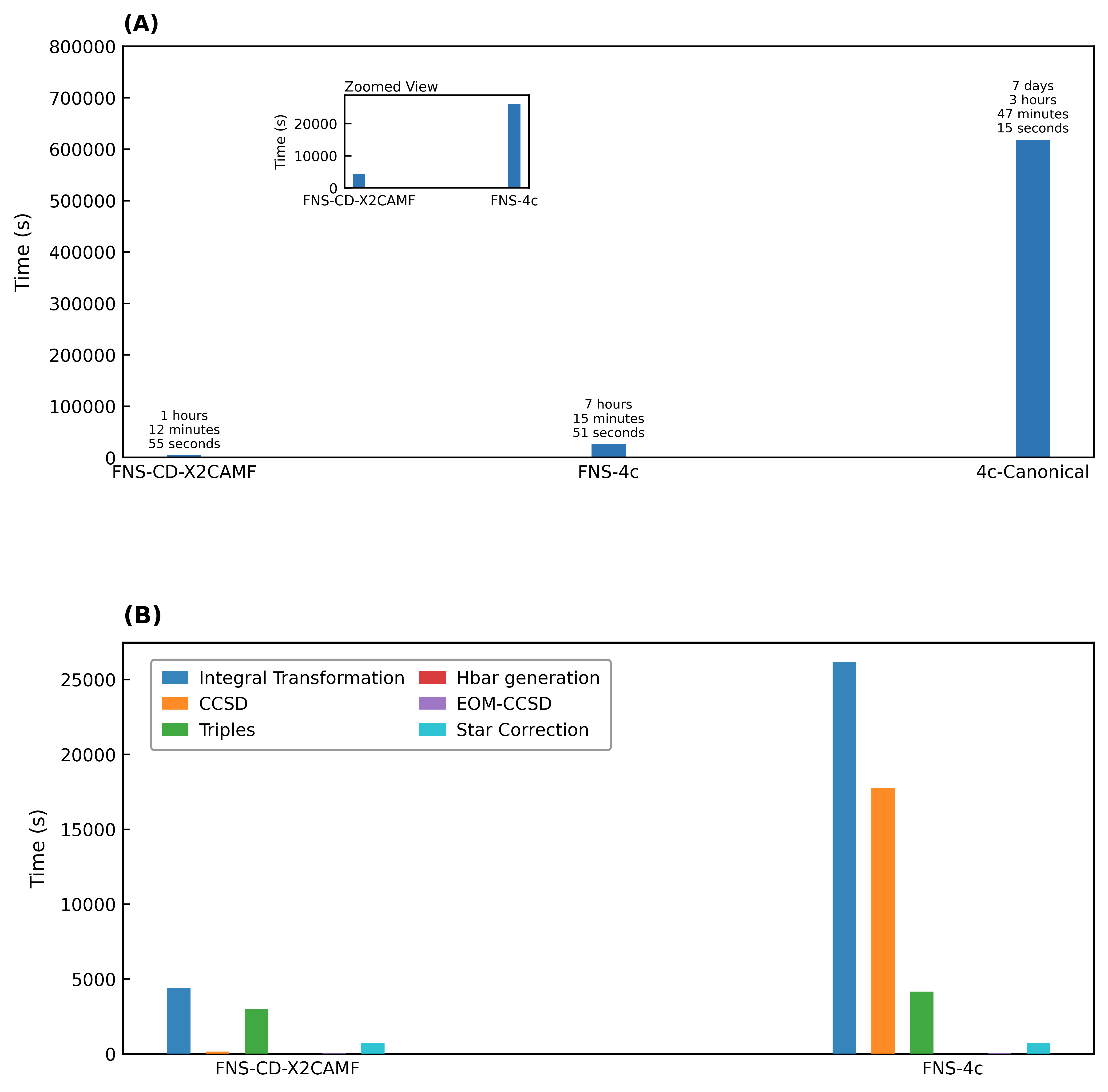}
    \caption{\justifying{(A) Comparison of wall times for the IP-EOM-CCSD(T)(a)$^\ast$ calculation using the 4c-canonical, 4c-FNS, and CD-X2CAMF-FNS implementations for the HI molecule. (B) Breakdown of individual timing components for the 4c-FNS and CD-X2CAMF-FNS implementations. The dyall.v4z basis set was used for both hydrogen and iodine atoms.}}
    \label{fig:time_cal}
\end{figure}

To evaluate the computational performance of the CD-based FNS-IP-EOM-CCSD(T)(a)$^\ast$ implementation within the X2CAMF framework, a comparative timing analysis was carried out for the HI molecule employing the canonical 4c, FNS-4c, and CD-based FNS-X2CAMF Hamiltonians. The dyall.v4z basis set was used for both hydrogen and iodine atoms, and the frozen-core approximation was applied throughout. The calculations were performed on a dedicated workstation equipped with two Intel Xeon Silver 4210R CPUs (2.40~GHz) and 512~GB of RAM. The canonical calculation comprised 18 occupied and 474 virtual spinors. As illustrated in Fig.~\ref{fig:time_cal}(A), the wall time of the IP-EOM-CCSD(T)(a)$^\ast$ calculation is substantially reduced in the FNS framework relative to the canonical 4c reference, primarily due to the reduced number of virtual spinors. Within the FNS framework, only 140 virtual spinors are retained. The canonical calculation required a total wall time of 7 days, 3 hours, 47 minutes, and 15 seconds, whereas the corresponding FNS calculation was completed in 7 hours, 15 minutes, and 51 seconds, corresponding to an approximate 24-fold reduction in computational time. Employing the CD-based X2CAMF Hamiltonian further reduces the computational cost dramatically, lowering the total wall time to 1 hour, 12 minutes and 55 seconds. This corresponds to an approximate 6-fold speedup relative to the FNS-4c calculation and about a 141-fold speedup compared to the canonical 4c calculation.

Fig.~\ref{fig:time_cal}(B) presents a detailed breakdown of the computational timings, comparing the CD-X2CAMF-FNS-IP-EOM-CCSD(T)(a)$^\ast$ and 4c-FNS-IP-EOM-CCSD(T)(a)$^\ast$ implementations. Consistent with our previous studies,\cite{CD_CCSD_2025, CD_IP_2025} the dominant reduction in total wall time arises from the substantially lower cost of generating two-electron integrals in the X2CAMF framework compared to the full 4c relativistic integral evaluation. The CD-X2CAMF approach also leads to a significant reduction in the time required for the CCSD ground-state calculation, in line with observations from earlier work.\cite{CD_CCSD_2025, CD_IP_2025} In contrast, the computational cost of the perturbative triples correction is only modestly reduced relative to the corresponding 4c calculations performed within the same FNS space. This limited speedup arises because several terms in both the ground-state and excited-state triples corrections involve exchange-type integrals, whose formal scaling cannot be reduced within the CD approximation employed in the present study. Nevertheless, the overall cost of the method is significantly reduced through the combined use of the FNS truncation and the CD-X2CAMF approximation.

\section{\label{sec:level5}Conclusion}
We have presented the theoretical formulation, implementation, and benchmarking of triples corrections to the relativistic IP-EOM-CC method. The full IP-EOM-CCSDT and several perturbative approximations to it were assessed on a representative set of test cases. Among all perturbative approximations studied, the IP-EOM-CCSD(T)(a)$^\ast$ method, which incorporates non-iterative triples corrections to both the ground and target states, provides the best compromise between computational cost and accuracy, and shows significant improvement over the relativistic IP-EOM-CCSD method. It avoids the storage of three-body amplitudes for both the ground and ionized states, thereby substantially reducing the memory footprint while preserving the leading-order effects of triple excitations. The method scales non-iteratively as $\mathcal{O}(N^7)$, with storage requirements comparable to those of the ground-state CCSD method. The number of floating-point operations is further reduced through the use of frozen natural spinors and the CD-X2CAMF approximation. An FNS threshold of $10^{-5}$ reproduces the results of the untruncated canonical calculation at a fraction of the computational cost. The CD-X2CAMF approximation yields results in close agreement with the full 4c DC Hamiltonian, with a significant reduction in both wall time and memory requirements. The newly developed FNS-X2CAMF-IP-EOM-CCSD(T)(a)$^\ast$ method can be routinely applied to atoms and small molecules, and opens the door to a wide range of applications in relativistic quantum chemistry and physics. Work towards such applications is currently in progress.

\section*{Supplementary Material}
The Supplementary Material contains the programmable expressions for the relativistic IP-EOM-CCSDT method.
\begin{acknowledgments}
The authors acknowledge support from IIT Bombay, the DST-SERB CRG grant (Project No. CRG/2023/002558), the IIT Bombay supercomputing facility, and C-DAC supercomputing resources (PARAM Smriti and PARAM Brahma). M.T. acknowledges DST for the INSPIRE fellowship. A.K.D. acknowledges the research fellowship funded by the EU NextGenerationEU program through the Recovery and Resilience Plan for Slovakia under project No. 09I03-03-V04-00117.
\end{acknowledgments}

\appendix
\section{Ground-State CCSD(T)(a) Triples Correction}
%%%%%%%%%%%%%%%%%% ENERGY DENOMINATORS %%%%%%%%%%%%%%%%%%%
{\small
\begin{equation} %%T1 denom
\label{eqn:D1}
e_{i}^{a}
= \varepsilon_{i} - \varepsilon_{a}.
\end{equation}
}

{\small
\begin{equation} %% T2 Denom
\label{eqn:D2}
e_{ij}^{ab}
= \varepsilon_i + \varepsilon_j - \varepsilon_a - \varepsilon_b.
\end{equation}
}

{\small
\begin{equation}  %% T3 denom
\label{eqn:D3}
(\hat{D}_{3})_{ijk}^{abc}
= \varepsilon_i + \varepsilon_j + \varepsilon_k
  - \varepsilon_a - \varepsilon_b - \varepsilon_c.
\end{equation}
}
%%%%%%%%%%%%%%%%%%%%%%%%%%%%%%%%%%%%%%%%%%%%%%%%%%%%%%%%%%%%%%%
{\small
\begin{equation}
\label{eqn:t3_unnorm}
X_{ijk}^{abc} 
= \sum_d \langle bc \Vert dk \rangle\, (t_{2}^{\mathrm{CCSD}})_{ij}^{ad}
-
\sum_m \langle cm \Vert kj \rangle\, (t_{2}^{\mathrm{CCSD}})_{im}^{ab}.
\end{equation}
}

{\small
\begin{equation}
\label{eqn:T3_2}
t_{3}^{[2]}
=
\frac{
\mathcal{P}(i,j,k)\,\mathcal{P}(a,b,c)\; X_{ijk}^{abc}
}{
(\hat{D}_{3})_{ijk}^{abc}
}.
\end{equation}
}

\subsection{Corrected Cluster Amplitudes}
%%%%%%%%%%%%%%%%%%%%%%%%%%%%%
% T1-DELTA
%%%%%%%%%%%%%%%%%%%%%%%%%%%%%

{\small
\begin{equation}
\label{eqn:T1delta}
t_{1}^{\Delta}
=
\frac{1}{4\, e_{i}^{a}}
\sum_{j k b c}
\langle jk \Vert bc \rangle\; (t_{3}^{[2]})_{ijk}^{abc}.
\end{equation}
}

%%%%%%%%%%%%%%%%%%%%%%%%%%%%%
% T2-DELTA
%%%%%%%%%%%%%%%%%%%%%%%%%%%%%
{\small
\begin{equation}
\label{eqn:T2delta}
\begin{aligned}
t_{2}^{\Delta}
&=
\frac{1}{e_{ij}^{ab}}
\Bigg[
\sum_{k c}
(t_{3}^{[2]})_{ijk}^{abc} f_{k}^{c}
\\[6pt]
&\quad
+ \frac{1}{2}
\Bigg(
\mathcal{P}(ab)
\sum_{k c e}
(t_{3}^{[2]})_{ijk}^{a e c}\,
\langle k b \Vert c e \rangle
\\[6pt]
&\qquad
+ \mathcal{P}(ij)
\sum_{k l c}
(t_{3}^{[2]})_{i l k}^{a b c}\,
\langle l k \Vert j c \rangle
\Bigg)
\Bigg].
\end{aligned}
\end{equation}
}
%%%%%%%%%%%%%%%%%%%%%%%%%%%%%%%%%%%%%%%%%%%%%%%%%%%%%%%%%%%%%%%%%%%
% FINAL CCSD(T)(a) AMPLITUDES
{\small
\begin{equation}
\label{eqn:T1corr}
t_{1}^{\mathrm{CCSD(T)(a)}} = t_{1}^{\mathrm{CCSD}} + t_{1}^{\Delta};
\end{equation}

\begin{equation}
\label{eqn:T2corr}
t_{2}^{\mathrm{CCSD(T)(a)}} = t_{2}^{\mathrm{CCSD}} + t_{2}^{\Delta}.
\end{equation}
}
%%%%%%%%%%%%%%%%%%%%%%%%%%%%%%%%%%%%%%%%%%%%%%%%%%%%%%%%%%%%%%%%%%%

% H-BAR INTERMEDIATE
\subsection{Corrected $\bar{H}$ Intermediate}
{\small
\begin{equation}
\label{eqn:Hbar_delta}
\bar{H}_{mcij}^{\Delta}
=
\frac{1}{2}
\sum_{k a b}
(t_{3}^{[2]})_{ijk}^{abc}\;
\langle m j \Vert a b \rangle .
\end{equation}
}
\section{Star Triples Correction for EOM}

\subsection{Left Vector: $[\widetilde{l}_{3}]_{ab}^{ijk}$}
\begin{equation}
\label{eqn:X1L}
X_{1}^{L}
=
\sum_{k}
\langle ij \Vert ab \rangle\, l^{k}
\end{equation}

\begin{equation}
\label{eqn:X2L}
X_{2}^{L}
=
\sum_{m}
\langle ji \Vert m a \rangle\;
l_{b}^{mk}
\end{equation}

\begin{equation}
\label{eqn:X3L}
X_{3}^{L}
=
\sum_{e}
\langle i e \Vert a b \rangle\;
l_{e}^{jk}
\end{equation}

\begin{equation}
\label{eqn:L3tilde}
\begin{aligned}
[\widetilde{l}_{3}]_{ab}^{ijk}
& =
\mathcal{P}(i,j,k)\!\left[X_{1}^{L}-X_{2}^{L}\right]
+ \mathcal{P}(a,b)\,X_{2}^{L}
+ \mathcal{P}(i,j,k)\,X_{3}^{L}.
\end{aligned}
\end{equation}

% ============================
\subsection{Right Vector: $[\widetilde{r}_{3}]_{ijk}^{ab}$}
\begin{equation}
\label{eqn:X1R}
X_{1}^{R}
=
\sum_{mn}
\langle mn \Vert jk \rangle\;
r_{n}\;
t_{i m}^{a b}
\end{equation}

\begin{equation}
\label{eqn:X2R}
X_{2}^{R}
=
\sum_{e}
\langle b a \Vert e i \rangle\;
r_{jk}^{e}
\end{equation}

\begin{equation}
\label{eqn:X3R}
X_{3}^{R}
=
\sum_{m,e}
\langle m b \Vert k e \rangle\;
r_{m}\;
t_{ij}^{a e}
\end{equation}

\begin{equation}
\label{eqn:X4R}
X_{4}^{R}
=
\sum_{m,b}
\langle a m \Vert i j \rangle\;
r_{m k}^{b}
\end{equation}

\begin{equation}
\label{eqn:R3tilde}
\begin{aligned}
\relax[\widetilde{r}_{3}]_{ijk}^{ab}
&=
\mathcal{P}(i,j,k)\!\left[X_{1}^{R}+X_{2}^{R}\right]
-
\mathcal{P}(a,b)\,\mathcal{P}(i,j,k)\!\left[X_{3}^{R}+X_{4}^{R}\right]
\end{aligned}
\end{equation}

\vspace{8mm}

\textbf{Permutation Notation}

{\small
\begin{flalign}
\mathcal{P}(ij)\, f(ij)
&=
f(ij) - f(ji), &
\\[6pt]
\mathcal{P}(ab)\, f(ab)
&=
f(ab) - f(ba), &
\\[8pt]
\mathcal{P}(ijk)\, f(ijk)
&=
f(ijk)
+
f(jki)
+
f(kij), &
\\[6pt]
\mathcal{P}(abc)\, f(abc)
&=
f(abc)
+
f(bca)
+
f(cab). &
\end{flalign}
}

%\nocite{*}
\bibliographystyle{aipnum4-1}
\bibliography{paper}

\end{document}